\documentclass[final,5p,times,twocolumn]{elsarticle}
\usepackage{amssymb}
\journal{}
\begin{document}
\begin{frontmatter}

\title{Wave-particle duality using the Compton effect}

\author[1,2]{Lucas S. Pollyceno}
\author[1]{Alexandre D. Ribeiro \corref{cor1}}
\address[1]{Universidade Federal do Paran\'a,
          Departamento de F\'{\i}sica, 
          Curitiba-PR,
          81531-990, 
          Brazil}
\address[2]{Universidade Estadual de Campinas, 
          Instituto de F\'{\i}sica “Gleb Wataghin,”
          Campinas-SP,
          13083-859, 
          Brazil}

\cortext[cor1]{corresponding author}
\ead{aribeiro@fisica.ufpr.br}

\begin{abstract}
 Thought experiments based on the double-slit interferometer had a crucial role to develop ideas concerning the wave-particle duality and the Bohr's complementarity principle. Ideally, a slit with a sufficiently low mass recoils due to the passage of the photon. This motion denounces the path taken by the light and suppresses any attempt to observe an interference pattern. In real life, however, available which-way information in such a setup is significantly impaired by the typical magnitudes of photons and slits, making the verification of the effect almost impossible. Here, we extend this discussion by applying similar ideas to the Mach-Zehnder interferometer. That is, we study the consequences of the beam-splitter recoil, during the passage of the photon, over the interference pattern produced by the device. Unlike the double-slit experiment, this recoil can now be encoded in the wavelength of the photon itself, which, in principle, is more easily accessed. Fortuitously, the model used to describe the interaction between the idealized beam-splitter and the photon clearly indicates that an interferometer based on Compton's effect could be build to study wave-particle duality. We follow this hint, finding realistic experimental parameters needed to observe the trade-off between wave and corpuscular behaviors in such a modified interferometer.
\end{abstract}


\end{frontmatter}

\section{Introduction}
\label{intro}

At the end of the XIX century, scientists had a very well established notion indicating that light behaved as a wave. Such a solid thought was built during a long time. In fact, almost a hundred years had been passed since the last paradigm shift involving this subject, when the observation of diffraction and interference phenomena was decisive for reprobating Newton's corpuscular theory of light, prevailing by that time. Crucial experiments performed in the beginning of the XX century, concerning, for instance, black-body radiation, photoelectric effect, and Compton scattering, unsettled that scenario because their satisfactory theoretical descriptions necessarily involved the hypothesis that light behaved as particles. Given the evident conflict, the term wave-particle duality was coined to refer to this coexistence of classically antagonistic behaviors: while wave properties are fundamental to understand some light phenomena, particle features are necessary to describe others. Besides, wave and particle behaviors were not simultaneously verified in the same experimental realization, an observation which is formalized by the Bohr's complementarity principle.

Quantum Mechanics, as it stands nowadays, accommodates this duality in a reasonable way. However, during the first years of the theory development, this issue was hotly debated by Albert Einstein and Niels Bohr~\cite{EinsteinBohr}. Essentially, while Bohr defended his complementarity principle, Einstein confronted him with a series of thought experiments where both manifestations could be brought out. Taking the emblematic double-slit experiment as an example, Bohr's complementarity basically implies that the interference pattern --- a wave signature --- cannot be observed together with the measurement of the slit by the which the photon travels --- a particle signature. It agrees with our classical physics sense because interference demands superposition of at least two sources (paths, in this case) of light. That is, if one observes the photon traveling through a given slit, there is no reason to get interference. Nevertheless, arguing that the interaction between photons and slits could produce a possibly detectable recoil of the latter in a sufficiently accurate experimental setup, Einstein challenged this principle. In fact, such a recoil could denounce, by means of the imparted momentum, the path traveled by the photon without disturbing the interference pattern ({\it a priori}), producing a counterexample for Bohr's statement. 

In his reply, Bohr used the Heisenberg uncertainty relation to successfully argue that it is impossible to determine the slit that the light passes through without creating an uncertainty on its location. As the interference pattern is sensitive to the slits position, one concludes that the fringes would be eliminated by the which-way measurement, recovering what is stated in Bohr's principle. In 1979, Wootters and Zurek~\cite{Zurek} reexamined this discussion in detail, favoring Bohr's argument. Their novelty, however, was to introduce an information-theoretic approach to this problem, in which the correlation between slit and photon, generated by their interaction, was announced as a fundamental quantity to understand the issue. Around a decade later, in a paper authored by Scully, Englert and Walther~\cite{scully1991}, using an alternative strategy which avoids the uncertainty relation argument, an experiment was proposed where Einstein's goal is achieved. In this new proposal, they deal with atom interferometry instead of light, and the crucial point is to reveal the path traveled by the atom using a cavity placed along each path. By means of the interaction between the atom and the field inside the cavity, which-way information can be obtained and one cannot effectively appeal to the uncertainty relation to explain the expected loss of the interference fringes. Interestingly, as suspected by Wootters and Zurek~\cite{Zurek}, the fundamental mechanism supporting the principle of complementarity was definitively changed to be the entanglement between the interfering object and the physical system responding by the which-way information. It is worth emphasizing that these ideas stimulated other possibilities to observe the traveled path, overcoming the almost impossible task to measure the recoil of a macroscopic object (a slit), when interacting with a microscopic particle (an atom or a photon).

Given this scenario, a number of works on duality and complementarity based on similar approaches have been developed~\cite{eichman,wiseman,rempe,haroche2001,lapuma,walborn}. In particular, in some of them~\cite{lapuma,walborn}, the role of the path informer is curiously assumed by an internal degree of freedom of the interfering photon itself (i.e., its polarization). More recently, provided by fascinating experimental advances, realizations much more similar to Einstein's proposal have been performed~\cite{Schmidt2013,Martin2015,Liu2015,teutlein,brand}, where a physical recoil of the slits really happens. Indeed, these works have somehow measured the momentum transferred from the scattered particle to the slits. Basically, their strategies consists of replacing slits by controlled molecules~\cite{Schmidt2013,Martin2015,Liu2015}, or making an array of single-layer graphene nanoribbons take the role of a diffraction grating~\cite{teutlein,brand}. All these events reported so far reveal that the great scientific interest attracted by the subject persists for almost one century. For completeness, we should mention that other important contributions around the theme were developed as, for instance, a necessary formalism in order to quantify wave and corpuscular behaviors \cite{englert,2015FP,qureshi} or a study involving a possible application to general relativity~\cite{arndt,margalit}.

\begin{figure*}[!t]
\centerline{
\includegraphics[width=13cm,angle=0]{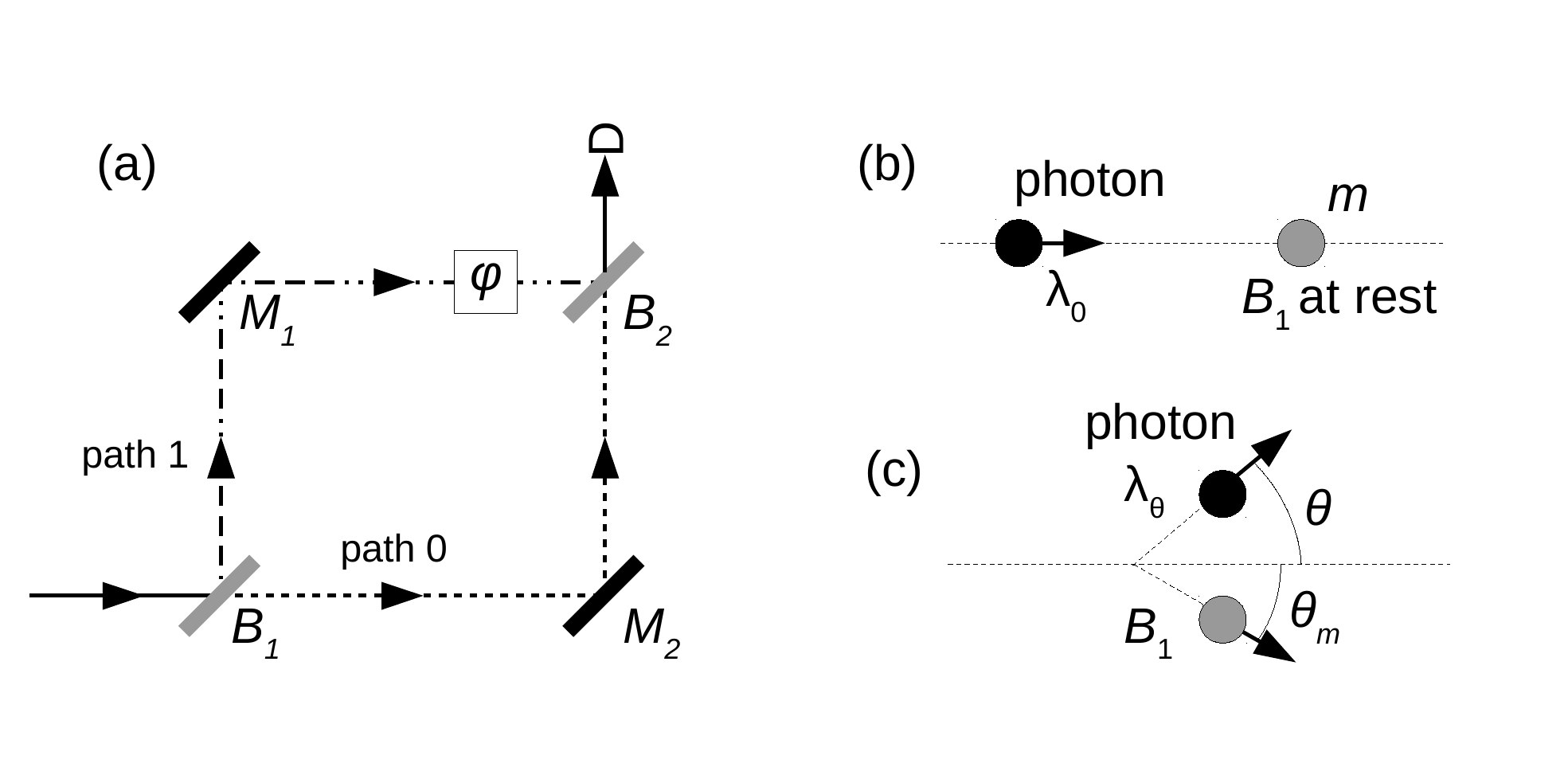}}
\caption{(a) Mach-Zehnder interferometer. A light beam (represented by the solid line) is sent to a beam-splitter $B_1$, dividing it into two others (dotted and dot-dashed lines) with the same intensity. Each beam travels different paths; one passing through the mirror $M_1$, and the other by $M_2$. As a consequence, right before the beam-splitter $B_2$, it is assigned a phase $\varphi$ between the beams. They are recombined by $B_2$ and measured by a detector $D$. Panels (b) and (c) illustrate a model where $B_1$ is replaced by a single particle of mass $m$. Before absorbing the photon [panel (b)], the particle is assumed to be at rest. After emitting the photon [panel (c)], considering energy and momentum conservation, the particle may receive a recoil in the direction $\theta_m$. Eventually, the emitted photon changes its direction with respect to the initial situation, as represented by the angle $\theta$ [for the interferometer of panel (a), $\theta$ should be 0 or $\pi/2$, exclusively].}
\label{fig1}
\end{figure*}

In the present paper, we study wave-particle duality using a Mach-Zehnder interferometer [Fig.~\ref{fig1}(a)], which essentially manifests the same phenomena as those seen in the double-slit device. In the figure, the interfering quantum system (light in our case) is sent to the interferometer, reaching the beam-splitter $B_1$; the two possible different paths after $B_1$ are represented by the dotted (path 0) and dot-dashed (path 1) lines. Then, after receiving a relative phase $\varphi$, the two paths are recombined by the beam-splitter $B_2$. A detector $D$ placed at path 0 observes the light obtained from this process. Clearly, in the case of a laser beam entering in the device, the intensity detected by $D$ is a function of $\varphi$, oscillating between zero and a certain maximum value. Such an interference pattern attests the wave behavior of the light. On the other hand, performing the experiment with photons individually sent to the device, and following Einstein's ideas~\cite{EinsteinBohr}, one questions about the possibility of getting which-way information from the interaction between the photon and $B_1$. In order to explore this point and check the equivalence with the double-slit experiment, we use an artificial but pedagogical model, where $B_1$ is assumed as a single particle initially at rest, which absorbs the photon, and immediately re-emits it, as illustrated in Fig.~\ref{fig1}(b) and Fig.~\ref{fig1}(c). To faithfully represent the Mach-Zehnder interferometer, the photon should be emitted into only two directions with the same probability: either $\theta=0$ or $\theta=\pi/2$ [for future convenience, in Fig.~\ref{fig1}(c) we represent a generic angle $\theta$]. In principle, the announced equivalence is clear: considering that energy and momentum are conserved during the interaction, if we were able to  measure the recoil of $B_1$, then we would obviously get path information for the photon and the annihilation of the interference pattern should be observed. A last comment should be done: except for the restriction concerning the values assumed by $\theta$, this model of photon-particle interaction is exactly the same used to explain the photon-electron interaction in the Compton effect~\cite{compton}.

There is no conceptual novelty in the discussion presented in the last paragraph. It is entirely contained in the debates of Einstein and Bohr~\cite{EinsteinBohr}. Including the similarities with the Compton work are identified there. We already know that, by measuring the imparted momentum of $B_1$, its position would get uncertain, causing the loss of the interference pattern. But we remind that the crucial point to not experimentally explore this idea is the lack of accuracy of measuring the recoil of a macroscopic object, due to the interaction with a microscopic one, in a realistic interferometer. Here enters our contribution. Compton taught us in his model that, when the photon deviates, the target particle necessarily recoils, and this information is embedded in the photon wavelength, which could be considered as one of its internal degrees of freedom. We intend, therefore, to develop a theory in order to evaluate the possibility of using the photon wavelength as an indirect measurement of the $B_1$ recoil, and, consequently, as a which-way informer. In this sense, our study resembles the approaches~\cite{lapuma,walborn} where photon polarization codifies which-way information. Notice that the use of these ideas in the double-slit interferometer discussed by Einstein and Bohr~\cite{EinsteinBohr} does not directly apply to because both paths in this case generate a similar changing in the photon wavelength. Although the above approach can be considered artificial since the physical beam-splitter is taken as a single particle, we intend to explore the fact that this interaction model is exactly the same used in the Compton scattering, where the role of the particle is played by a free electron. As a result, we will determine some parameters for a Compton-effect-based interferometry where wave-particle duality can be observed. Curiously, almost one hundred years after its appearance to consolidate corpuscular aspects of light, here we develop a way to use Compton's model to also reveal its wave behavior.

We organize the paper as follows. In Sect.~\ref{model} we present the artificial Mach-Zehnder interferometer considering that its first beam-splitter is a single particle. Then, in Sect.~\ref{compton}, we show the connection between this description and the Compton scattering, finding a more realistic model to determine the parameters needed to observe the two regimes manifested by the light. Our final remarks are presented in Sect.~\ref{FR}.

%%%%%%%%%%%%%%%%%%%%%%%%%%%%%%%%%%%%
\section{Idealized beam-splitter model}
\label{model}

In this section we develop the idea of assuming the beam-splitter $B_1$ of Fig.~\ref{fig1}(a) as a single particle of mass $m$ ($B_2$ continues as an ordinary beam-splitter). In this case, a photon with wavelength $\lambda_0$ enters in the interferometer, and is absorbed by $B_1$, initially at rest [Fig.~\ref{fig1}(b)]. Then, it immediately emits the photon along two possible directions with equal probabilities [Fig.~\ref{fig1}(c), with $\theta=0$ or $\theta=\pi/2$]. If the photon is sent to path 1 (dot-dashed line; $\theta=\pi/2$), the process is equivalent to a reflection under the action of $B_1$. If it is emitted in path 0 (dotted line; $\theta=0$), we say that it simply passed through the beam-splitter.

Concerning the interaction process between the photon and $B_1$, we will assume that relativistic energy and momentum are conserved. Initially, energy and momentum of the photon are, respectively,
\begin{equation}
  E^{(i)}_p = \frac{hc}{\lambda_0}
  \quad\mathrm{and}\quad
  \vec{p}^{(i)}_p = \frac{h}{\lambda_0} \hat{k},
\end{equation}
where $\hat{k}$ is a unitary vector denoting the horizontal direction (to the right), $h$ is the Planck constant, and $c$ is the speed of light in vacuum. In addition, at the initial time, $B_1$ has null momentum and energy $E^{(i)}_{m} = mc^2$. After the emission, the energy and momentum of the photon are given by
\begin{equation}
E^{(f)}_p = \frac{hc}{\lambda_\theta}
\quad\mathrm{and}\quad
\vec{p}^{(f)}_p = \frac{h}{\lambda_\theta} \hat{{\theta}},
\end{equation}
respectively, where $\hat{{\theta}}$ denotes the direction and $\lambda_\theta$ the wavelength of the emitted photon. The beam-splitter energy and momentum are, at the final time,
\begin{equation}
E^{(f)}_{m} =\sqrt{p^2_{m}c^2+m^2c^4}
\quad\mathrm{and}\quad
\vec{p}_{m} = {p}_{m} \hat{{\theta}}_m,
\end{equation}
respectively, where $\hat{{\theta}}_m$ is  unitary vector denoting the direction into the which $B_1$ recoils. Clearly, according to momentum conservation, if the initial direction of the photon propagation is changed ($\theta=\pi/2$), then $B_1$ necessarily recoils, as illustrated in Fig.~\ref{fig1}(c). Even though the direct measurement of the motion of $B_1$ be impracticable, we emphasize that it is also encoded in the wavelength of the scattered photon, 
\begin{equation}
  \lambda_\theta = \lambda_0 + \frac{h}{m c}(1-\cos\theta)
  \quad\Longrightarrow\quad
  \lambda_{\frac{\pi}{2}} = \lambda_0 + \frac{h}{m c},
  \label{lambdatheta}
\end{equation}
result which is achieved by manipulating the conservation laws. Returning to the Mach-Zehnder interferometer, we conclude that, after the beam-splitter $B_2$, when the paths are recombined to reach $D$, path information is contained in the photon itself. Naturally, the question concerning the real possibility of distinguishing $\lambda_0$ and $\lambda_{\frac{\pi}{2}}$ arises at this point, because the wavelength difference is very small,
\begin{equation}
  \delta\lambda \equiv \lambda_{\frac{\pi}{2}}-\lambda_0 =\frac{h}{mc}\approx
  2.4~\mathrm{pm}.
  \label{deltaL}
\end{equation}
However, we will discuss this point later, when we examine the feasibility of the model as a whole.

\begin{figure*}[!t]
\centerline{
\includegraphics[width=8cm,angle=0]{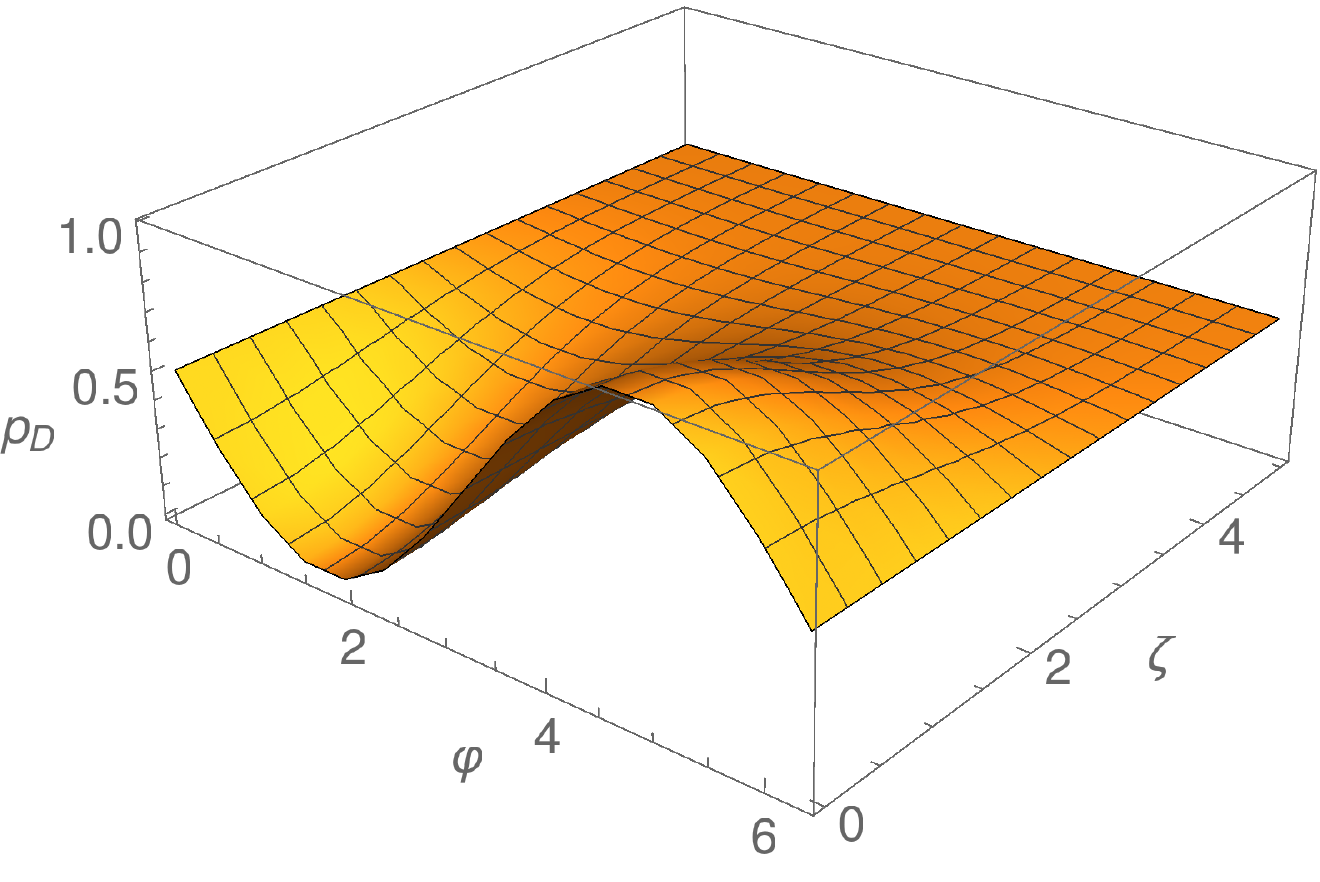}}
\caption{Probability $p_{_D}$ of a photon to reach detector $D$ as function of the relative phase $\varphi$ and the ratio $\zeta\equiv\delta\lambda/\sigma$. For small values of $\zeta$, function $p_{_D}$ oscillates with maximum amplitude by varying $\varphi$; in the opposite limit, $p_{_D}$ is a flat function, insensitive to $\varphi$.}
\label{fig2}
\end{figure*}

So far the interaction model between photon and beam-splitter was presented. Now we proceed with the description of the device shown in Fig.~\ref{fig1}(a). The (pure) state of the photon entering in the interferometer is denoted by $|\psi_0\rangle = |0,\lambda_0\rangle$, where the label 0 is included to refer to its initial path of propagation (path 0). After the idealized beam-splitter, according to the above theory, we consider that the state of the photon is given by the superposition
\begin{equation}
  |\psi_1\rangle = \frac{|0,\lambda_0\rangle +
    i|1,\lambda_{\frac{\pi}{2}}\rangle }{\sqrt2},
  \label{psi1}
\end{equation}
meaning that the photon can go up (path 1) with wavelength $\lambda_{\frac{\pi}{2}}$ or continue to right (path 0) keeping its original wavelength. Notice that a phase $e^{i\pi/2}=i$ due to the reflection is assigned to the second ket. More importantly, the state $|\psi_1\rangle$ is entangled: degrees of freedom related to the {\em path taken} and {\em wavelength} are now correlated. Following the same reasoning and including the phase $\varphi$ gained by traveling in path 1, we conclude that the state right after $B_2$ is 
\begin{equation}
  |\psi_2\rangle = \frac{1}{\sqrt2}\left[
    \frac{|0,\lambda_0\rangle+i|1,\lambda_0\rangle
    }{\sqrt2}+e^{i\varphi}
    \frac{i|0,\lambda_{\frac{\pi}{2}}\rangle+
      |1,\lambda_{\frac{\pi}{2}}\rangle}{\sqrt2}
  \right].
\end{equation}
Given the fact that the detector does not measure photon wavelengths, the theoretical description should define a density operator $\rho_2 \equiv |\psi_2\rangle\langle\psi_2|$ and take the partial trace over the degree of freedom corresponding to the photon wavelength. Assuming that $|\lambda_0\rangle$ and $|\lambda_{\frac{\pi}{2}}\rangle$ may not be orthogonal to each other, i.e., that $|\lambda_0\rangle$ and $|\lambda_{\frac{\pi}{2}}\rangle$ do not represent strictly  monochromatic states, the reduced density operator in the path basis $\{|0\rangle,|1\rangle\}$ becomes
\begin{equation}
  \rho_2^{\mathrm{red}} = \frac12\left(\begin{array}{cc}
    1- |A|\sin(\varphi+\delta) & |A| \cos(\varphi+\delta) \\
    |A| \cos(\varphi+\delta) & 1+ |A|\sin(\varphi+\delta)
  \end{array}\right).
  \label{rhored}
\end{equation}
Here, we also define $\langle\lambda_0 | \lambda_{\frac{\pi}{2}}\rangle \equiv A= |A|e^{i\delta}$ and assume that $\langle\lambda_0 | \lambda_{0}\rangle= \langle\lambda_{\frac{\pi}{2}} | \lambda_{\frac{\pi}{2}}\rangle=1$. By projecting $\rho_2^{\mathrm{red}}$ to path 0, where the detector is placed, we find the probability of the photon be measured by $D$,
\begin{equation}
  p_{_{D}} =\frac12\left[1- |A|\sin(\varphi+\delta)\right] .
  \label{pD}
\end{equation}

Clearly, when the two wavelengths are totally distinguishable, $| \lambda_{0}\rangle$ is orthogonal to $| \lambda_{\frac{\pi}{2}}\rangle$ and $A=0$, implying that $p_{_{D}}\to \frac12$. That is, the photon entering in the interferometer has 50\% chance of being detected in $D$. Obviously, it has the same probability to be detected in another detector, if placed in path 1, after $B_2$. Notice that, for the case $A=0$, clicks in the detector $D$ do not depend on $\varphi$, indicating corpuscular behavior, as expected since which-way information is available, even if it is not accessed. In the opposite case, when the two wavelengths are completely undistinguishable, $A=1$ and the probability of a click be observed on $D$ depends on the phase $\varphi$: $p_{_{D}}\to \sin^2\left[\frac12\left(\varphi-\frac{\pi}{2}\right) \right]$, attesting the wave behavior. Again, the lack of wich-way information retains the observation of an interference pattern.

To illustrate the intermediate case where $0<|A|<1$, it is convenient to consider a model in which
\begin{equation}
  A\equiv\langle\lambda_0 | \lambda_{\frac{\pi}{2}}\rangle =
  \int_0^\infty  \Lambda_0(\lambda)~\Lambda_{\frac{\pi}{2}}(\lambda)
~\mathrm{d}\lambda.
\label{Aint}
\end{equation}
The wavelength distribution $\Lambda_{\xi}(\lambda)$ depends on a number of factors that affect the light source, such as quantum uncertainties in the emission process (homogeneous line broadening), the Doppler effect due to the motion of the emitter, and interaction with the environment (inhomogeneous line broadening)~\cite{demtroder}. As our intention is just to illustrate the phenomenon, we will simply consider the wavelength distribution as a normalized Gaussian centered at $\lambda_\xi$, with variance $\sigma^2$,
\begin{equation}
\Lambda_{\xi}(\lambda) = \left(\frac{1}{\sigma\sqrt{\pi}}\right)^{1/2}
\exp\left[ -\frac{1}{2\sigma^2} (\lambda-\lambda_\xi)^2\right].
\label{wldistr}
\end{equation}
From this expression, provided that $\lambda_0$ and $\lambda_\frac{\pi}{2}$ are sufficiently far from $\lambda=0$, we can straightforwardly evaluate $A=|A|e^{i\delta}$, finding 
\begin{equation}
  \delta=0
  \quad\mathrm{and}\quad
  |A|=\exp{\left[-\frac{1}{4}
      \left(\frac{\delta\lambda}{\sigma}\right)^2\right]},
\end{equation}
where $\delta\lambda$ is given by Eq.~(\ref{deltaL}). Notice that the ratio $\zeta\equiv\delta\lambda/\sigma$ is a measure for the distinguishability between the states $| \lambda_{0}\rangle$ and $| \lambda_{\frac{\pi}{2}}\rangle$; broader and closer distributions imply small values of $\zeta$ and indistinguishable states, while sharper and distant ones imply large $\zeta$ and distinguishable states. In Fig.~\ref{fig2}, we illustrate the trade-off between wave and particle behaviors. Notice that, when $\zeta\equiv\delta\lambda/\sigma$ is negligible, the states $| \lambda_{0}\rangle$ and $| \lambda_{\frac{\pi}{2}}\rangle$ are indistinguishable, which means absence of which-way information. Function $p_{_D}$, therefore, is $\varphi$-dependent as light were a wave. On the other hand, when $\zeta\equiv\delta\lambda/\sigma$ is large ($\zeta\approx5$ in Fig.~\ref{fig2}), the two states are effectively orthogonal, so that information about the path traveled is available. In this case, there is no dependence on $\varphi$, a manifestation of a corpuscular behavior.

Clearly, the inclusion of the electron state in the present description does not change the conclusions presented above. It can be easily verified by replacing $|0,\lambda_0\rangle\to|0,\lambda_0,\vec{p}_m^{\,(0)}\rangle$ and $|1,\lambda_\frac{\pi}{2}\rangle\to|0,\lambda_\frac{\pi}{2},\vec{p}_m^{\,(\pi/2)}\rangle$ in Eq.~(\ref{psi1}), where $\vec{p}_m^{\,(\theta)}$ refers to the electron recoil when the photon is emmited in the $\theta$ direction. Then, following the calculation, and taking the trace also over the $\vec{p}_m^{\,(\theta)}$ variable, because it is also unaccessible, we recover the same reduced operator of Eq.~(\ref{rhored}), from the which our conclusions were deduced.

The physical system presented in this section indicates how the photon wavelength could be used as a which-way informer. However, we still did not discuss the possibility to observe these predictions in a real experiment, which is the task of the next section.

%%%%%%%%%%%%%%%%%%%%%%%%%%%%%%%%%%%%%%%%%%%%%%
\section{Compton-scattering-assisted duality}
\label{compton}

\begin{figure*}[!t]
\centerline{
  \includegraphics[width=5.3cm,angle=0]{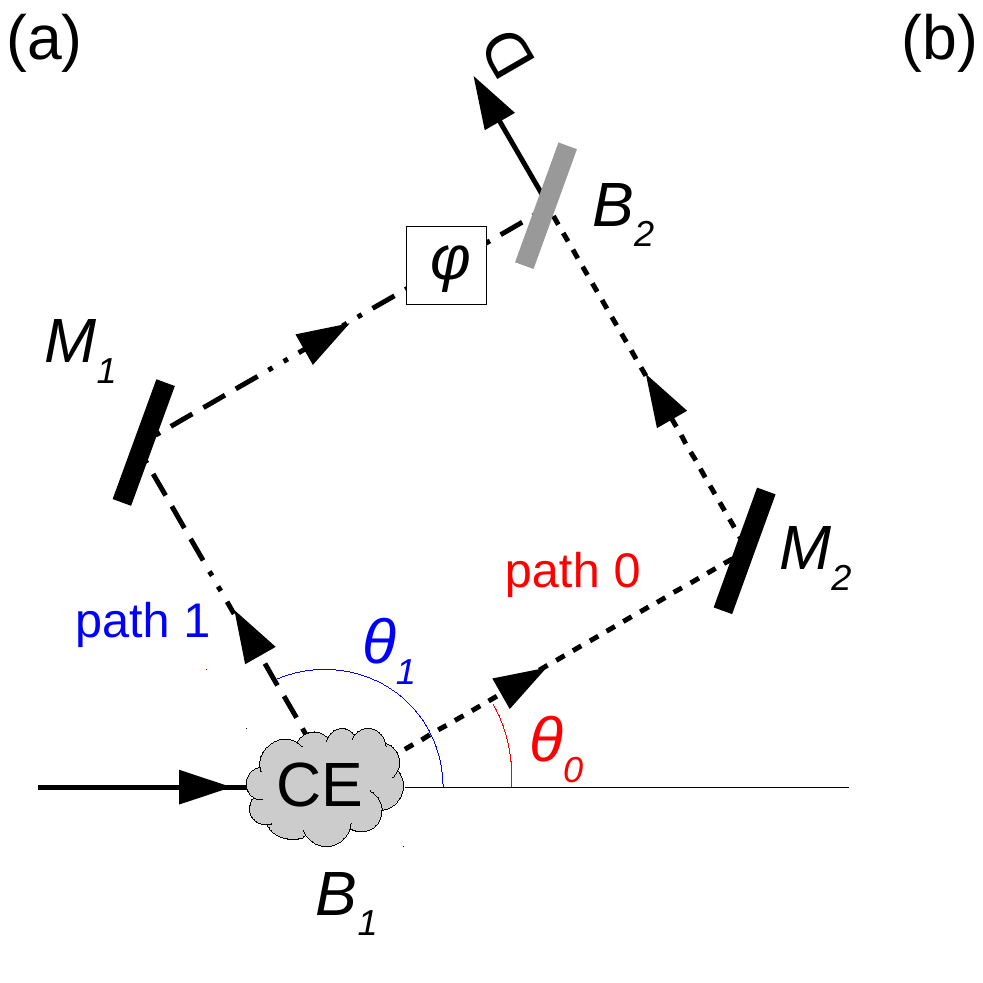}\hspace{-.5cm}
  \includegraphics[width=7.5cm,angle=0]{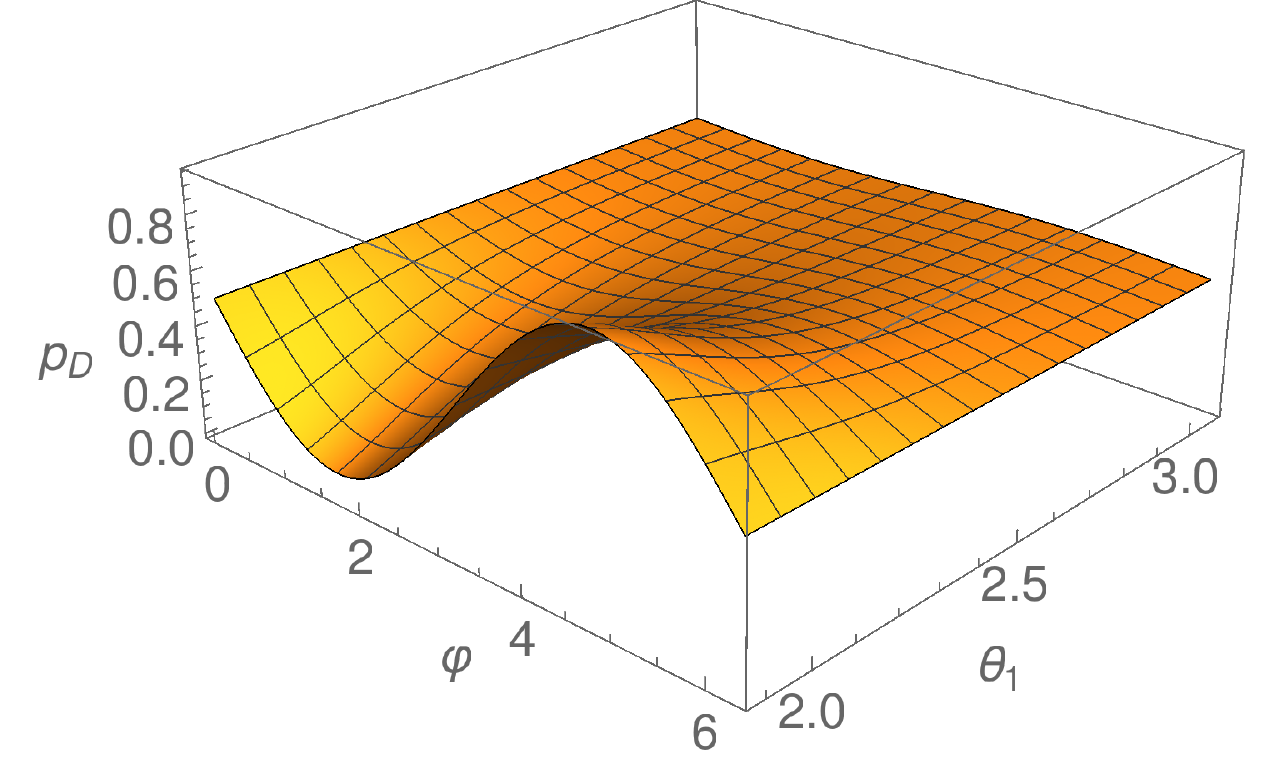}}
\caption{(a) A modified setup for the Mach-Zehnder interferometer. A light beam (represented by the solid line) reaches the device $B_1$, and suffers an inelastic scattering modeled according to the Compton effect (CE). Two scattering angles $\theta_0$ and $\theta_1$ are selected to compose the paths 0 and 1 of the original Mach-Zehnder setup [Fig.~\ref{fig1}(a)]. After $B_1$, the modified interferometer is similar to the original one. In panel (b), for the interferometer of panel (a), we show the probability $p_{_D}$ of a photon reaches detector $D$ as function of the relative phase $\varphi$ and $\theta_1$. The value of $\theta_0$, for each $\theta_1$, is determined by the condition of equiprobability, expressed by the black solid line of Fig.~\ref{fig3}(d). While wave behavior ($\varphi$-dependence) is manifested when $\theta_1<2$, particle behavior ($\varphi$-independence) appears for $\theta_1>3$.}
\label{fig4}
\end{figure*}

An updated view of Compton's work~\cite{compton}, supported by quantum theory, indicates that the state of the photon, after being scattered by the electron, consists of a continuous superposition of states relative to all possible scattering directions $\theta$. Besides, each term of this summation represents a state where propagation direction and wavelength $\lambda_\theta$ are correlated. That is,
\begin{equation}
  |\mathrm{scattered\;\;state}\rangle = \int c(\theta)
  |\theta,\lambda_\theta\rangle~\mathrm{d}\theta.
  \label{scatterstate}
\end{equation}
Therefore, in order to fit the ideas presented in the last section to a more realistic scenario, we will argue that we can use a device drawn to observe the Compton effect and select two distinct scattering directions ($\theta_0$ and $\theta_1$) to constitute paths 0 and 1 of a Mach-Zehnder interferometer [see Fig.~\ref{fig4}(a)]. In our theoretical description, it basically means that we should apply the projector $|\theta_0\rangle\langle \theta_0|+|\theta_1\rangle\langle \theta_1|$ to Eq.~(\ref{scatterstate}), getting a state similar to $|\psi_1\rangle$ of Eq.~(\ref{psi1}). However, to guarantee a correct equivalence between the two approaches after passing through $B_1$, our first task is to choose equiprobable directions, i.e., we should have $c(\theta_0)=c(\theta_1)$. By doing so, a minor difference between $|\psi_1\rangle$ and the resulting state remains: the phase gained by one of the paths due to reflection in the original setup. However, notice that this discrepancy can be ignored because it can be included in the relative phase~$\varphi$.

To find the equiprobable angles, we can look at the differential cross section of the scattered (unpolarized) photons derived by Klein and Nishina~\cite{KN,weinberg}, and obtained via quantum electrodynamics,
\begin{equation}
  \frac{d\sigma(\theta)}{d\Omega} =
  \frac{r_0^2}{2} \left(\frac{\lambda_0}{\lambda_\theta}\right)^2
  \left[\frac{\lambda_0}{\lambda_\theta}+
  \frac{\lambda_\theta}{\lambda_0}-
  \sin^2\theta\right],
  \label{crosssection}
\end{equation}
where $\lambda_{\theta}$ and $\lambda_{0}$ are expressed in Eq.~(\ref{lambdatheta}), $r_0=e^2/(4\pi\epsilon_0~mc^2)$ is the classical electron radius, $\theta$ is the polar angle in spherical coordinates, and $d\Omega$ is an infinitesimal element of solid angle. Notice that the notation adopted here is consistent with the description of the last section; the incident photon is assumed to travel along the $z$-direction, and $\theta$ refers to the deviation from this path. Here, the term $\lambda_\theta/\lambda_0$ also respects Eq.~(\ref{lambdatheta}), written as
\begin{equation}
  \frac{\lambda_\theta}{\lambda_0} =
  1+\epsilon(1-\cos\theta),
\end{equation}
and depends exclusively on the angle $\theta$ and the dimensionless parameter $\epsilon\equiv(hc/\lambda_0)/(mc^2)$, which is the ratio between the energy of the incident photon and the electronic rest energy.

It should be noticed that the deduction of Eq.~(\ref{crosssection}) is performed in the framework of the Quantum Field Theory, using a reference system where the electron is initially at rest. As argued in Weinberg's book~\cite{weinberg}, this is a reasonable assumption because electrons move non-relativiscally in atoms, while the described phenomenon is essentially relativistic. A huge advantage arises by considering this point. It states that there exists a plausible scenario where the thermal behavior of electrons can be neglected, fact which is also confirmed by Compton in his experiment. Therefore, in the present paper, we work with this hypothesis, and we do not deal with a thermal description for electrons.

\begin{figure*}[!t]
\centerline{
  \includegraphics[width=5.9cm,angle=0]{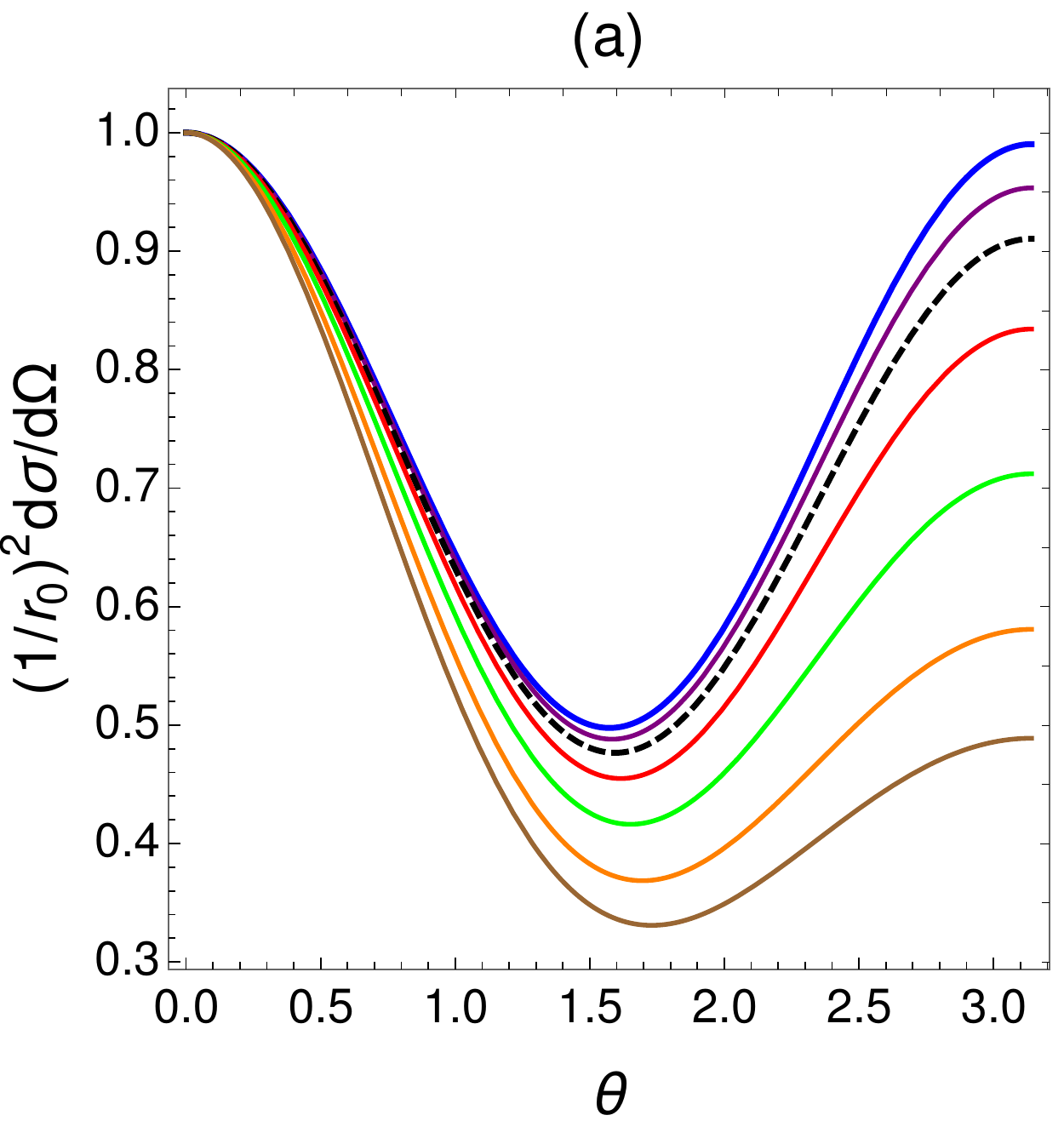}
  \includegraphics[width=5.9cm,angle=0]{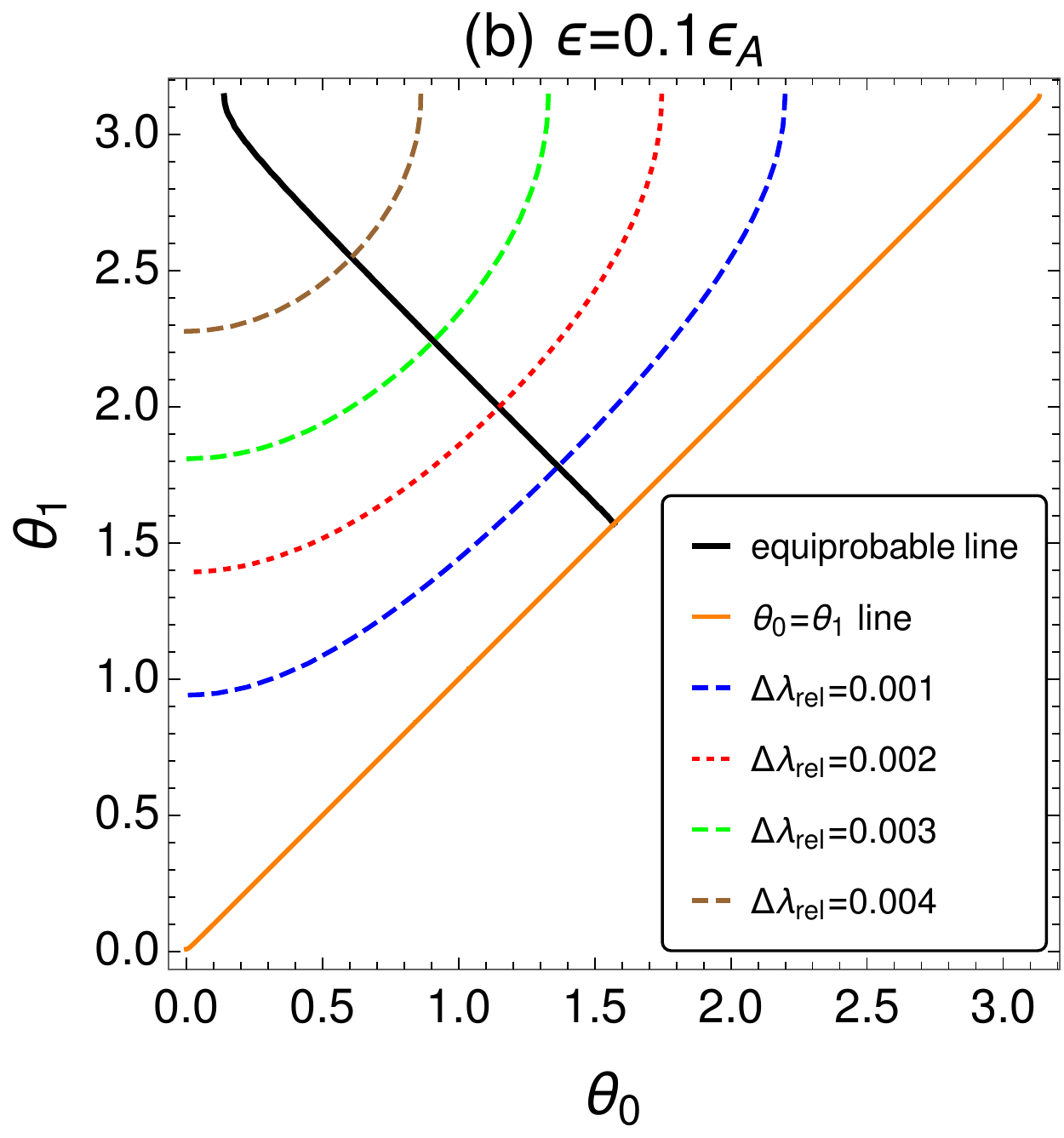}}
\centerline{
  \includegraphics[width=5.9cm,angle=0]{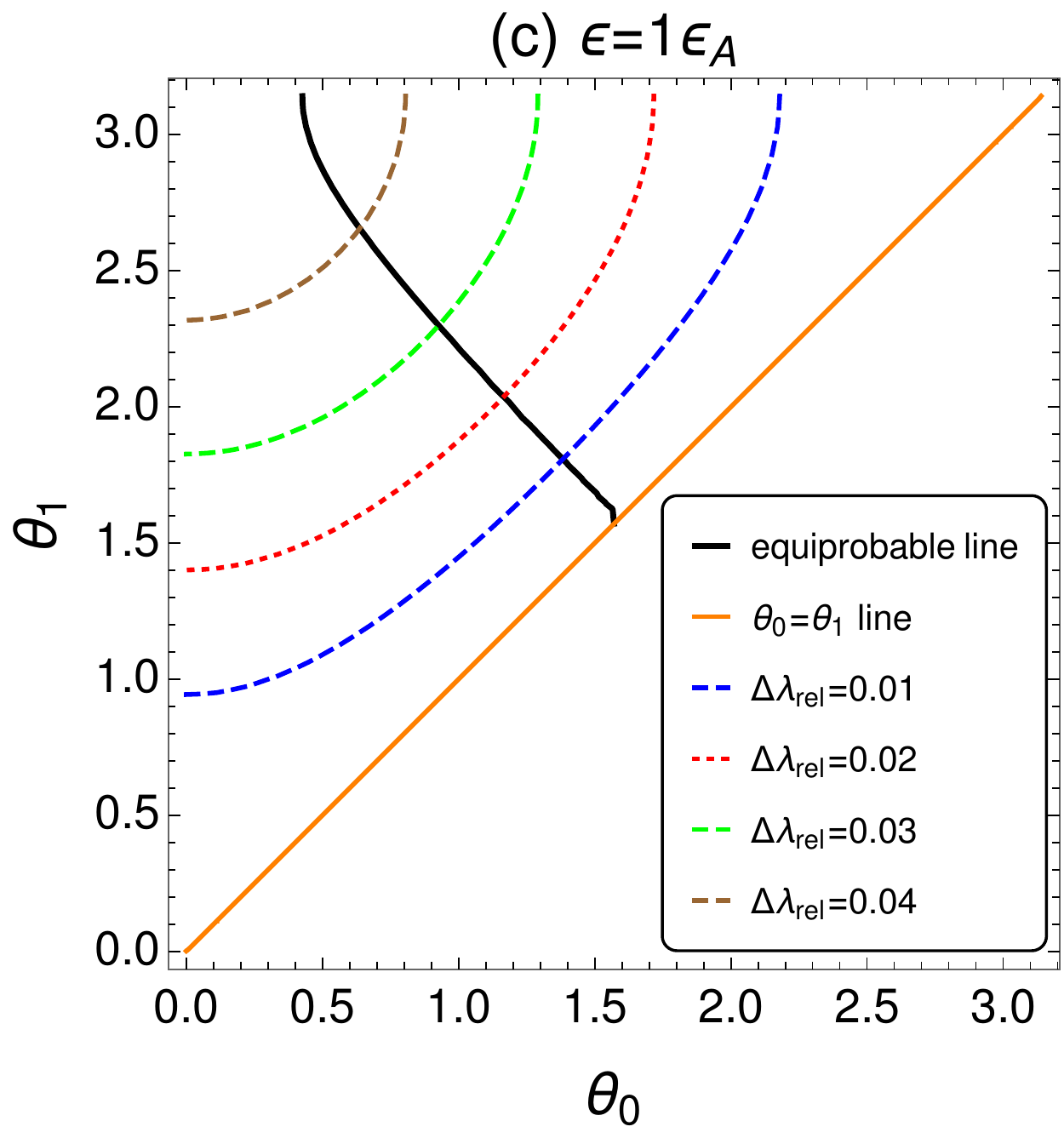}
  \includegraphics[width=5.9cm,angle=0]{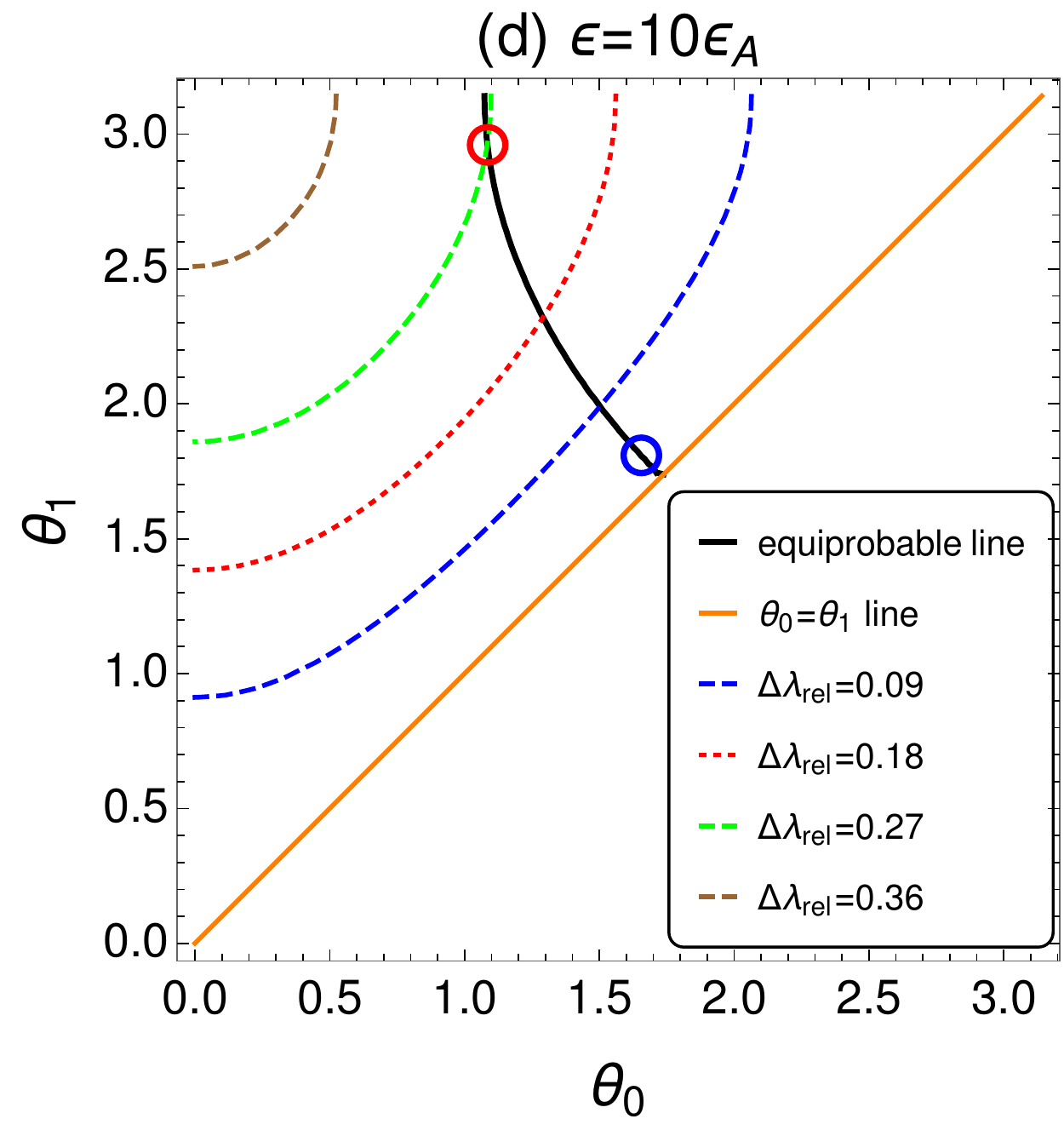}}
\caption{(a) The dimensionless differential cross section $(1/r_0^2)(d\sigma/d\Omega)$ as function of $\theta$. From the lowest to the highest curve, we have $\epsilon=10\epsilon_{_\mathrm{A}}$, $\epsilon=7\epsilon_{_\mathrm{A}}$, $\epsilon=4\epsilon_{_\mathrm{A}}$, $\epsilon=2\epsilon_{_\mathrm{A}}$, $\epsilon=1\epsilon_{_\mathrm{A}}$ (black dashed line), $\epsilon=0.5\epsilon_{_\mathrm{A}}$, and $\epsilon=0.1\epsilon_{_\mathrm{A}}$, with $\epsilon_{_\mathrm{A}}\approx0.0243$. Panel (b) shows, in the $(\theta_0,\theta_1)$-plane, for $\epsilon=0.1\epsilon_{_\mathrm{A}}$, the equiprobable curve (black solid line) and some contours where $\Delta \lambda_{\mathrm{rel}}$ is constant. As all these curves are invariant when $\theta_0$ and $\theta_1$ are interchanged, we present only results in the region $\theta_1>\theta_0$. Panels (c) and (d) are equivalent to panel (b), for $\epsilon=1\epsilon_{_\mathrm{A}}$ and $\epsilon=10\epsilon_{_\mathrm{A}}$, respectively.}
\label{fig3}
\end{figure*}

In Fig.~\ref{fig3}(a), we illustrate the behavior of the dimensionless differential cross section $(1/r_0^2)(d\sigma/d\Omega)$ as function of the scattering angle $\theta$, for some values of $\epsilon$: from the lowest to the highest curve, we have $\epsilon=10\epsilon_{_\mathrm{A}}$, $\epsilon=7\epsilon_{_\mathrm{A}}$, $\epsilon=4\epsilon_{_\mathrm{A}}$, $\epsilon=2\epsilon_{_\mathrm{A}}$, $\epsilon=1\epsilon_{_\mathrm{A}}$ (dashed line), $\epsilon=0.5\epsilon_{_\mathrm{A}}$, and $\epsilon=0.1\epsilon_{_\mathrm{A}}$. Here, $\epsilon_{_\mathrm{A}}$ is chosen as the factor $\epsilon$ for the case of an incident photon whose wavelength is 1\AA, which amounts to $\epsilon_{_\mathrm{A}}\approx0.0243$. Such a reference number was taken because, in typical experiments, we have $\lambda_0\sim 1$\AA. Clearly, for any value of $\epsilon$ shown in the figure, one can find two angles $\theta_0$ and $\theta_1$ that correspond to the same scattering probability density. However, these values are not given by 0 and $\pi/2$ as required by the original interferometer, implying that both path 0 and path 1 should be really changed to other directions.

In order to help us in the decision about which pairs $(\theta_0,\theta_1)$ would be better suited for an experiment, in Fig.~\ref{fig3}(b), using $\epsilon=0.1\epsilon_{_\mathrm{A}}$, we plot, in the $(\theta_0,\theta_1)$-plane, the equiprobable curve (black solid line), composed by the points where
\begin{equation}
  \frac{d\sigma(\theta_0)}{d\Omega}=
  \frac{d\sigma(\theta_1)}{d\Omega}.
\end{equation}
Were the purpose solely to choose equiprobable angles, then any point over the black solid curve of Fig.~\ref{fig3}(b) could be used in the device, provided that the energy of the incident photon was such that $\epsilon=0.1\epsilon_{_\mathrm{A}}$. However, in our study, it is also important to evaluate the distinguishability between the wavelengths $\lambda_{\theta_0}$ and $\lambda_{\theta_1}$, assigned to each direction. To this end, we define the relative difference as 
\begin{equation}
  \Delta \lambda_{\mathrm{rel}} \equiv
  \frac{|\lambda_{\theta_1}-\lambda_{\theta_0}|}
       {\frac12\left(\lambda_{\theta_1}+\lambda_{\theta_0}\right)},
\end{equation}
and plot some of its contour lines superimposed in Fig.~\ref{fig3}(b). To define a simple criterion concerning the experimental capability of discriminating two close wavelengths, we calculate the relative difference between $\lambda_{0}$ and $\lambda_{\frac{\pi}{2}}$ [see Eq.~(\ref{lambdatheta})], with $\epsilon=\epsilon_{_\mathrm{A}}$, getting $\Delta \lambda_{\mathrm{rel}}^{\mathrm{ref}}\equiv0.0240$. These parameters resemble those adopted in the original Compton experiment, so that we can infer that a typical device can distinguish two wavelengths, $\lambda_{\theta_0}$ and $\lambda_{\theta_1}$, for which $\Delta \lambda_{\mathrm{rel}} \gtrsim \Delta \lambda_{\mathrm{rel}}^{\mathrm{ref}}$. We should clarify that the action of distinguishing two wavelengths, at this point of the discussion, means to be able to identify their two distinct peaks in the spectrum, which is independent of any assumption concerning spectral widths. 

Notice that, in Fig.~\ref{fig3}(b), there is no pair $(\theta_0,\theta_1)$ over the equiprobable curve whose relative difference satisfies $\Delta \lambda_{\mathrm{rel}}\gtrsim \Delta \lambda_{\mathrm{rel}}^{\mathrm{ref}}$. Then, according to the adopted criterion of distinguishability, we need to move to other experimental parameters, since the role of path-informer, which is expected to be played by the wavelengths, risks to be hidden in this configuration. This is done in Fig.~\ref{fig3}(c) and \ref{fig3}(d), which show results equivalent to Fig.~\ref{fig3}(b), but for $\epsilon=1\epsilon_{_\mathrm{A}}$ and $\epsilon=10\epsilon_{_\mathrm{A}}$, respectively. Finally, in both figures, we can clearly find pairs of equiprobable directions for which the distinguishability criterion is respected. Although Fig.~\ref{fig3}(c) also presents potentially appropriate parameters to study wave-particle duality, from now on, we will restrict ourselves to the case in which the expected effect tends be more expressive. This is the situation contemplated by Fig.~\ref{fig3}(d), where the incident light is such that $\epsilon=10\epsilon_{_\mathrm{A}}$.

Returning to the Mach-Zehnder interferometer, now assisted by the Compton effect, we suggest the following experimental setup in order to testify the trade-off between corpuscular and wave behaviors of the light. We can first select the pair of angles $(\theta_0,\theta_1)$ as $(\theta_0^{(i)},\theta_1^{(i)})=(1.075,3.072)=(61.59^\circ,176.01^\circ)$ to indicate the new directions of paths 0 and 1, as shown in Fig.~\ref{fig4}(a). This pair refers to the point of Fig.~\ref{fig3}(d) highlighted by a red circle, where the relative difference is $\Delta \lambda_{\mathrm{rel}}^{(i)}=0.27$. Given that these parameters imply peak-distinguishability between their respective $\lambda_{\theta_0}^{(i)}$ and $\lambda_{\theta_1}^{(i)}$, measurements on the detector $D$ are expected to denounce particle behavior, that is, the intensity of light reaching $D$ should be insensitive to the relative phase $\varphi$. As we will show soon, even considering a spectral width of 10\% of $\lambda_0$, this behavior is still predicted. Starting from $(\theta_0^{(i)},\theta_1^{(i)})$, we can conclude that an option to continuously recover the wave behavior consists of changing the angles $\theta_0$ and $\theta_1$, still restricted to the equiprobable line of Fig.~\ref{fig3}(d), but going to the direction of $(\theta_0^{(f)},\theta_1^{(f)})=(1.590,1.882)=(91.10^\circ,107.83^\circ)$, which is the point marked by the blue circle. At these coordinates, we have $\Delta \lambda_{\mathrm{rel}}^{(f)}=0.01$, which may indicate indistinguishable $\lambda_{\theta_0}^{(f)}$ and $\lambda_{\theta_1}^{(f)}$, implying lack of path-information and the resurgence of the interference pattern.

In Fig.~\ref{fig4}(b), we illustrate the results obtained from all this analysis. The probability $p_{_{D}}$ [see Eq.~(\ref{pD})] of a photon reaching the detector $D$ is plotted as function of the relative phase $\varphi$ and also the angle $\theta_1$. The value of $\theta_0$, for each $\theta_1$, is determined by the equiprobability condition, and, therefore, can be found using the black solid line of Fig.~\ref{fig3}(d). We emphasize that, to calculate $p_{_{D}}$, a model for the wavelength distribution should be considered. Here, we also use the Gaussian function of Eq.~(\ref{wldistr}), with $\sigma=0.1\lambda_0$. Concerning the incident photon, to be consistent with Fig.~\ref{fig3}(d), we keep $\epsilon=10\epsilon_{_{\mathrm A}}$. Using these numerical parameters, in Fig.~\ref{fig4}(b), it is shown that the trade-off between particle and wave behaviors can be observed in this modified experimental setup of the Mach-Zehnder interferometer. The interferometer arms, when configured according to the point marked by the blue circle of Fig.~\ref{fig3}(d), give rise to a result where ``waviness'' is manifested ($\varphi$-dependence). It can be seen in Fig.~\ref{fig4}(b) for $\theta_1<2$. Contrarily, when the arms are determined by red circle of Fig.~\ref{fig3}(d), the $\varphi$-dependence is practically suppressed, revealing the particle behavior. This regime is manifested in Fig.~\ref{fig4}(b) for $\theta_1>3$. By comparing Fig.~\ref{fig2} and Fig.~\ref{fig4}(b), it should be pointed out that, while in the former plot $p_{_{D}}$ effectively oscillates between 0 and 1, in the latter, the amplitude of the oscillation is smaller. Also, the equivalent of the flat limiting surface seen Fig.~\ref{fig2} (for $\zeta\approx5$) now presents an almost undetectable oscillation. Essentially, these differences mean that, in this realistic configuration, we did not reach the situation where the two paths are completely distinguishable, neither the regime where they are totally indistinguishable. In spite of that, parameters adopted to plot Fig.~\ref{fig4}(b) are clearly appropriate to observe wave-particle duality in this Compton-based Mach-Zehnder interferometer.

\section{Final Remarks}
\label{FR}

In the present paper, we study wave-particle duality of light using an approach where the photon wavelength itself plays the role of a path-informer. We have conceptually developed this idea considering a Mach-Zehnder interferometer where the first beam-splitter is ideally assumed as a single particle. By identifying the interaction model between photon and beam-splitter as the same used do describe the Compton effect, we started to combine its physical description with that of the interferometer. Then, assuming realistic parameters, we proposed an original route to observe wave-particle duality in the resulting Compton-based Mach-Zehnder interferometer.

We adopted here a delicate but ordinary assumption, which is to neglect the electronic motion before its interaction with the photon. The absence of this property in an experiment clearly invalidates our description. However, according to Ref.~\cite{weinberg}, as we deal with a high-energy photon-electron scattering, where the electron is usually moving non-relativistically, it is a reasonable approximation to consider the electron initially at rest. Therefore, in the proposed experimental setup, we point out that this condition should be carefully studied and fulfilled.

Certainly some experimental difficulties were neglected in the present paper. For instance, it is well known that the Compton scattering simultaneously happens with an elastic scattering. That is, there are photons leaving $B_1$ with unchanged wavelength to all directions that could destroy our predictions. We believe that optical filters can be used to avoid their presence in the paths 0 and 1. In addition, choosing the arm angles $\theta_0$ and $\theta_1$ necessarily implies to consider a finite solid angle around the directions, which may demand some adjustments. In spite of these (and probably others) technical difficulties, it seems that an experiment could be performed to confirm the predictions. Interestingly, after almost one century from its publication, the Compton effect could be again involved in fundamental issues of quantum mechanics, but now with a different perspective.

At last, two comments deserve attention. First, leaving the Compton scattering aside for a while, we just speculate that an attempt to build low-massive beam-splitters could be done in the framework of optomechanical systems~\cite{OM}, similarly, for instance, to the idea of the quantum mirror treated in Ref.~\cite{saldanha}. In this case, the photon wavelength would also play the role of quantum-informer, according to the discussion of Sec.~\ref{model}. Second, it should be noticed that the connection between the two themes, duality and Compton scattering, have already been considered in another approach~\cite{pisk}. In that paper, cross sections of light scattered by different targets are calculated, and, when compared with certain patterns, the radiation behavior can be evaluated.

\section*{Acknowledgements}

We thank the following agencies for financial support: Conselho Nacional de Pesquisa\,--\,Brasil (CNPq); Coordena\c{c}\~ao de Aperfei\c{c}oamento de Pessoal de N\'{\i}vel Superior\,--\,Brasil (CAPES);  Instituto Nacional de Ci\^encia e Tecnologia - Informa\c{c}\~ao Qu\^antica\,--\,Brasil (INCT-IQ, 465469/2014-0); and Funda\c{c}\~ao Arauc\'aria. We also would like to thank R. M. Angelo, P. H. Souto Ribeiro, R. Medeiros de Ara\'ujo, I. A. Heisler, P. Milman, and R. Rabelo for their stimulating comments on this work. In particular, P. H. Souto Ribeiro should also be acknowledged for organizing the PROCAD meeting 2017, where fruitful discussions collaborated to develop this paper.\\

\end{document}